\documentclass[10pt,twocolumn,twoside,letterpaper]{IEEEtran}

\usepackage{geometry}
\makeatletter
\def\ps@IEEEtitlepagestyle{
  \def\@oddfoot{\mycopyrightnotice}
  \def\@evenfoot{}
}
\def\mycopyrightnotice{
  {\footnotesize
  \begin{minipage}{\textwidth}
  \centering
  978-1-7281-4164-0/19/\$31.00 \copyright2019 IEEE
  \end{minipage}
  }
}

\usepackage{url}
\usepackage{graphicx}
\usepackage{siunitx}

\begin{document}

\title{Operation and Calibration of a\\
Highly Granular Hadron Calorimeter\\ 
with SiPM-on-Tile Read-out
}
%
%
%

\author{O. Pinto for the CALICE Collaboration

\thanks{ 
Manuscript received December 13, 2019. This project has received funding from the European Union Horizon 2020 Research and Innovation programme under Grant Agreement no. 654168.

O. Pinto is with the Deutsches Elektronen-Synchrotron (DESY), Germany (e-mail: olin.pinto@desy.de)}}


\maketitle

\pagenumbering{gobble}

\begin{abstract}

The Analogue Hadron Calorimeter (AHCAL) is being developed within the CALICE collaboration for experiments at a future lepton collider. It is a sampling calorimeter with alternating layers of steel absorber plates and plastic scintillator tiles as active material. In the SiPM on tile design, the tiles are directly coupled to silicon photomultipliers (SiPMs). The front-end electronics are integrated into the active layers of the calorimeter. They are designed for low power consumption by rapidly power cycling according to the beam structure of a linear accelerator. In 2017 and 2018, a new large prototype with 38 active layers of 72x72 cm$^{2}$ size has been built. Each active layer consists of four readout boards with four 36 channel SPIROC2E readout ASIC each, resulting in 576 channels per layer. The prototype has been designed for mass production and assembly techniques: It uses injection-moulded tiles which were wrapped semi-automatically in reflector foil, tiles and electronics components were assembled using pick-and-place machines and all detector parts were tested during assembly. The  prototype with \SI {}{\sim}22,000 channels was commissioned at DESY and took muon, electron and pion data at the CERN SPS to demonstrate the capabilities of a SiPM-on-tile calorimeter concept with scalable detector design and the reliable operation of a large prototype. This proceedings gives an overview of the calibration results of the large CALICE AHCAL technological prototype.

\end{abstract}


\section{Introduction}
%
%
%
%
\IEEEPARstart{T}{he} physics at future high-energy lepton colliders demands a jet energy reconstruction with unprecedented precision. It can be reached by employing Particle Flow Algorithms (PFA)\cite{MAThom}, which are capable of achieving this by measuring each particle in the detector component with the best energy resolution. These algorithms  require  highly  granular  calorimeters  to  deliver optimal performance in jet energy resolution of the order of \SI{3}-\SI{4}{\percent} for jet energies between 45 GeV and several 100 GeV. Within the CALICE collaboration several highly granular technologies for electromagnetic and hadronic calorimeters are developed. One of these technologies is the Analog Hadron Calorimeter (AHCAL)\cite{Adloff}.

\section{The CALICE SiPM-on-Tile Hadron Calorimeter technological Prototype}

\begin{figure}[htp]
\centering
\includegraphics[width=8cm]{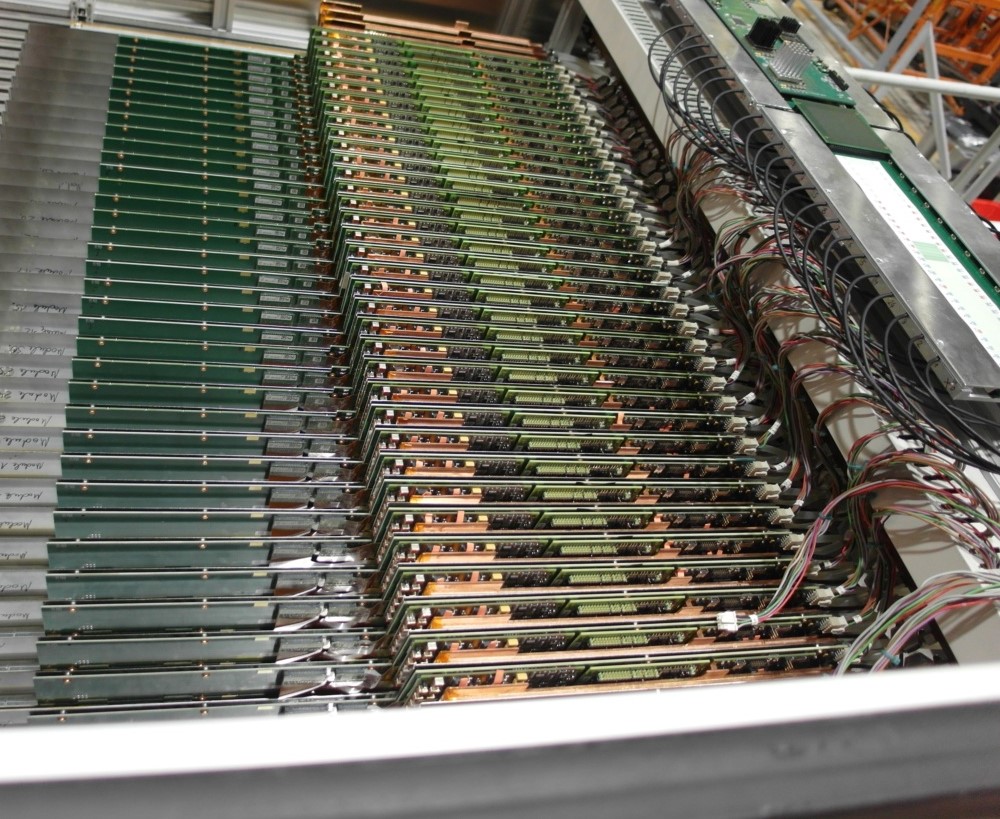}
\caption{Top view of AHCAL technological prototype fully assembled with 38 active layers}
\label{fig:Prot}
\end{figure}

\begin{figure}[htp]
\centering
\includegraphics[width=8cm]{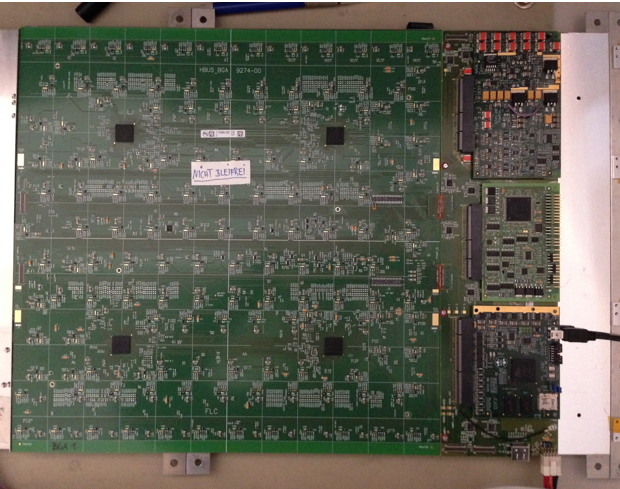}
\caption{HCAL Base Unit.}
\label{fig:hbu}
\end{figure}

The AHCAL technological prototype consists of a non-magnetic stainless steel absorber structure with approximately 17.2 mm thick plates and 38 active layers which are placed in the absorber structure as shown in Fig.\ref{fig:Prot}. It has 21,888 channels read out on an individual basis by SiPMs. The basic unit of the active elements is the HCAL Base Unit (HBU)\cite{Mat}, which has a size of 36x36 cm$^{2}$. It is equipped with 144 SiPMs which are read out by four SPIROC2E ASICs\cite{bou}. The key features of the electronics are its integration into the active volume and its capability to operate in power-pulsed mode which reduces the power consumption and eliminate the need for active cooling by exploiting the beam structure of linear colliders. The SPIROC2E ASIC also provides a cell-by-cell auto trigger and individual hit time with a precision of a few ns in test beam operations. In linear collider operation conditions with shorter data-taking windows, sub-ns time resolution is expected. The  prototype  is constructed with injection-moulded  polystyrene  scintillator tiles\cite{Liu} with a dimple at the centre optimised for best light collection\cite{liu}. They are read out by Hamamatsu  MPPC S13360-1325PE  photon  sensors. The technological prototype has a layout scalable to full a ILC detector.

\section{CALIBRATION}

The calibration procedure of the AHCAL prototype has several goals:

\begin{itemize}
\item The conversion of the hit energy and time from electronics to physical units
\item if necessary, the equalisation of the cell response to allow an efficient auto-triggering in all channels
\item the monitoring of the stability of the detector
\end{itemize}

A global calibration chain is necessary to properly calibrate the calorimeter channels. Every channel has to be calibrated separately due to non-uniformities such as unequal tile wrapping, glueing or SiPM and ASIC specific effects. Two calibration chains are needed, one for energy calibration and the other for time calibration. The hit energy or SiPM charge is read-out as ADC (Analogue to Digital Converter) and needs to be calibrated in units of minimum ionising particles (MIPs). The time is stored as a charge TDC (Time to Digital Converter) and is calibrated into values of nanoseconds. \\ The calibrated hit energy $E_{i}$ in units of MIP can be determined by eq. \ref{EnergyCalibration}:

\begin{equation}
E_{i}\hspace{0.15cm} [MIP] = f_{desat} \hspace{0.15cm} (pixels)  \frac{(A_{i}[ADC] - Pedestal)/IC}{MIP}
\label{EnergyCalibration}
\end{equation}
where
\begin{equation}
pixels = \frac{(A_{i}[ADC] - Pedestal)/IC}{Gain}
\label{pixel}
\end{equation}
the function f$_{desat}$ is the inverse of the SiPM response function, describing the SiPM output signal as a function of the incoming light intensity. At small amplitudes f$_{desat}$ is close to one and increases strongly for large signals.  IC is the inter-gain calibration.\\ As a final step in calibration, cluster energies will be converted to the GeV scale using electron testbeam data of known energies.

\subsection{Gain Calibration}

\begin{figure}[htp]
\centering
\includegraphics[width=8cm]{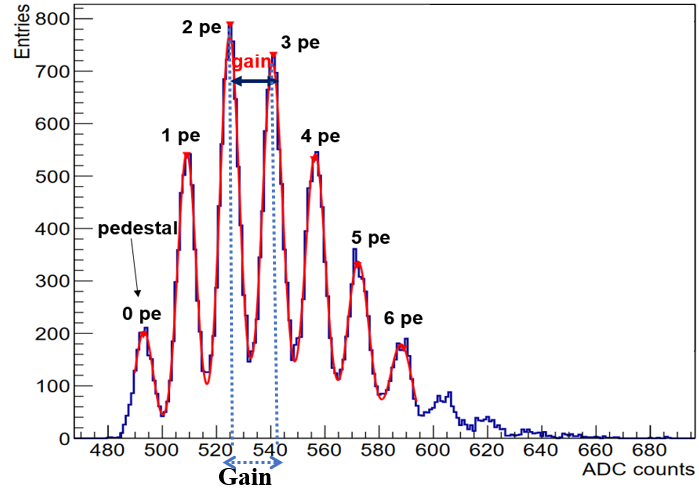}
\caption{SiPM single photoelectron spectrum of a single channel.}
\label{fig:SiPM_response}
\end{figure}

\begin{figure}[htp]
\centering
\includegraphics[width=8cm]{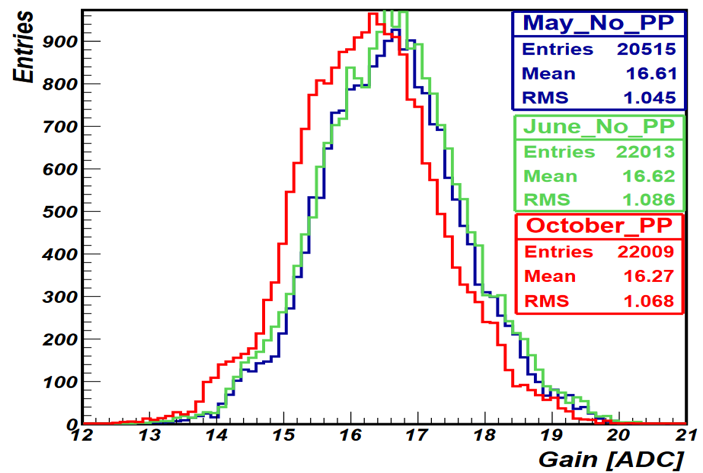}
\caption{Gain constant extracted for all AHCAL channels.}
\label{fig:Total_gain_calibration}
\end{figure}

The gain calibration allows the possibility of monitoring the detector performance and stability. The gain calibration is performed using dedicated LED runs. For all the AHCAL channels, single photon spectra are taken at low intensity LED light, and are fitted with a multi-Gaussian function. The gain value is extracted as the distance between the first and the second single photon-electron peaks. An example of a typical single photon spectrum is shown in Fig \ref{fig:SiPM_response}. The distance between two neighbouring peaks is left as a single free parameter in the multi-Gaussian fit. The width of the peaks is dominated by electronic noise. The statistical uncertainty on the gain determination is less than  \SI{1}{\percent} for fits which pass quality criteria. \\
During testbeam operation LED runs were taken approximately once per day. Gain extraction was achieved for \SI{95}{\percent} of the channels. The gain values for the remaining \SI{5}{\percent} of channels were taken as the average of the channels of the corresponding chip. Combining  several gain runs, more than \SI{99.9}{\percent} of all channels can be calibrated individually. The mean gain is ${\sim}$16.6 ADC/pixel with a spread of 1.0 ADC/pixel which is about \SI{6}{\percent} as shown in Fig \ref{fig:Total_gain_calibration}. For channels on the same ASIC the spread is only \SI{2.5}{\percent}, showing that the gain variation between ASICs contributes significantly to the overall spread. The gain is consistent between the May 2018 run and the June 2018 run (both without power pulsing). For the October 2018 run (with power pulsing) a small shift in gain is observed which was caused by electronic effects.

\subsection{MIP Calibration}

\begin{figure}[htp]
\centering
\includegraphics[width=8cm]{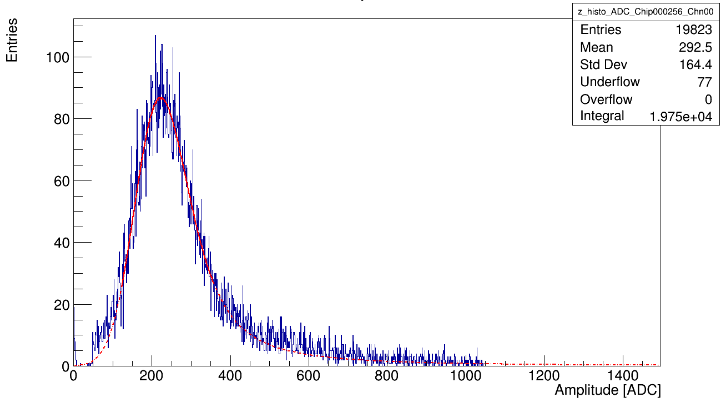}
\caption{Energy deposited by a muon in one AHCAL cell. The red solid line is obtained by fitting a Landau convoluted to a Gaussian function to the data. The dotted red line is the function extrapolated outside the fitting range. }
\label{fig:mip_spectrum}
\end{figure}

\begin{figure}[htp]
\centering
\includegraphics[width=8cm]{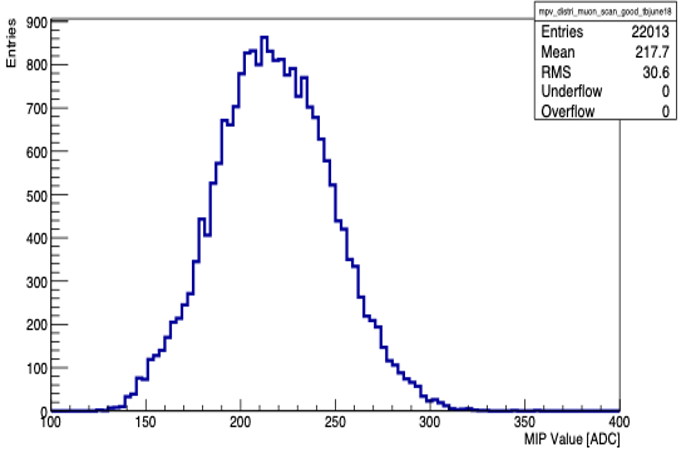}
\caption{MIP constant extracted for all AHCAL channels.}
\label{fig:total_mip}
\end{figure}

Minimum-ionising particles (MIPs) are used to perform a cell to cell equalization. For each channel the measured hit energy spectrum for muons is analysed. Its most probable value is derived from a fit with a Landau function convoluted with a Gaussian. An example of a single channel is shown in Fig \ref{fig:mip_spectrum}.
\\
The fit function takes into account not only the energy loss of muons within the scintillator tiles, but also contributions from photon counting statistics and electronic noise.  The MIP calibration efficiency, i.e. the amount of successful fits of muon hit energy spectra, is about \SI{99.92}{\percent}. As shown in Fig \ref{fig:total_mip} the mean MIP calibration constant is 217 ADC with an RMS spread of 30 ADC which corresponds to \SI {}{\sim}\SI{14}{\percent}. For channels on the same ASIC, the spread is \SI{8}{\percent}, which is good enough for autotrigger operation with the same threshold for all channels.

\subsection{Inter-Gain Calibration}
\begin{figure}[htp]
\centering
\includegraphics[width=8cm]{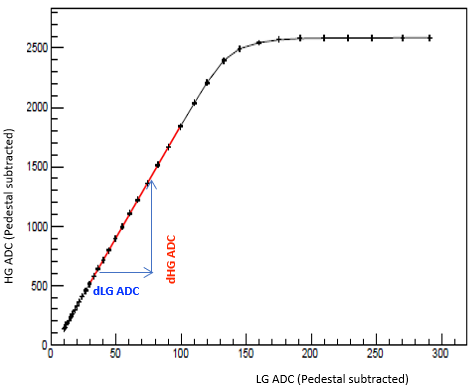}
\caption{Response of one AHCAL channel in the two amplifier gain modes (Low Gain and High Gain) for increasing LED light intensity.}
\label{fig:IC}
\end{figure}

The SPIROC2E chip can store the ADC either in low gain (LG) or in high gain (HG) mode. The inter-gain calibration factor of the channel is the ratio of HG to LG and is required for smooth transition between HG and LG hit energies. This constant depends on ratio of capacitors because of the circuit design. The extraction of the inter-gain calibration coefficients from LED runs is possible because the range of linear response of the chip in both modes overlaps. The amplitude of the signal is varied within the linear range of HG and LG by varying the LED light intensity. The inter-calibration coefficient for each channel is taken as the slope of a linear fit as shown in Fig \ref{fig:IC}. The inter-calibration factor ranges between 17 and 21 with a mean of 19.4 and a RMS spread of 0.66  which is \SI {}{\sim} \SI{3}{\percent}. For the remaining channels the inter calibration values were taken as the average of the channels of the corresponding chip.

\begin{figure}[htp]
\centering
\includegraphics[width=8cm]{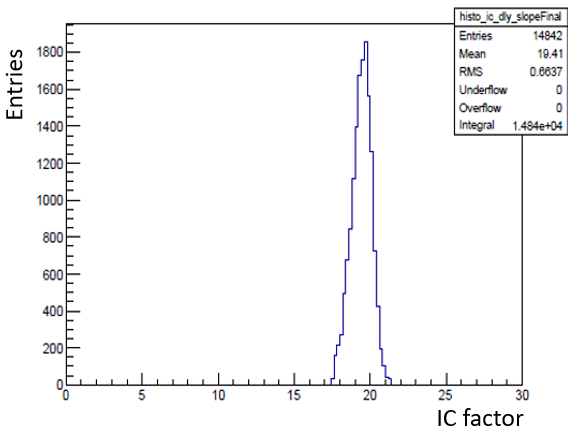}
\caption{Inter-gain calibration values extracted for all calibrated channels. }
\label{fig:IC_all}
\end{figure}

\subsection{SiPM Saturation}

\begin{figure}[htp]
\centering
\includegraphics[width=8cm]{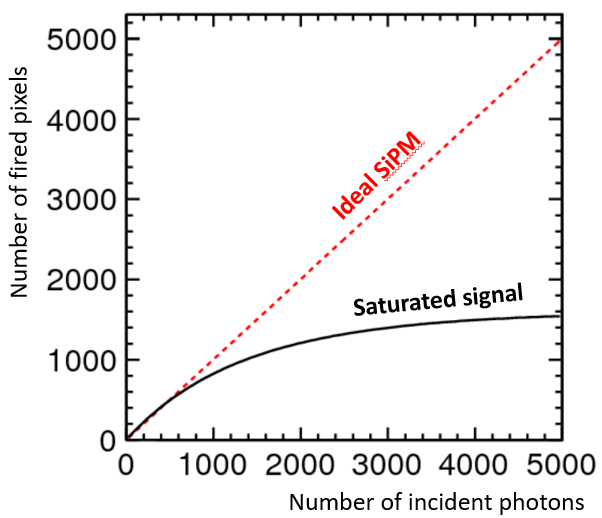}
\caption{Describing the response of SiPM.}
\label{fig:saturation}
\end{figure}

Due to the limited number of pixels and the finite pixel recovery time, the SiPM is an intrinsically a non-linear device and limited by response  saturation at high amplitudes. The response function of a SiPM describes the  number of fired pixels as a function of the number of incoming photons (see figure~\ref{fig:saturation}). It can be approximated by
\begin{equation}
N_{fired}^{sat} = N_{eff} \hspace{0.25cm}  (1 - e^{N_{fired}^{unsat}/N_{eff}}),
\end{equation}
where $N_{fired}^{sat}$ is the number of fired pixels and $N^{eff}$ is the effective number of pixel in the SiPM. 
For studies of saturation behaviour bare SiPMs were used. A dedicated set-up was used to ensure that all the pixels were illuminated in a homogeneous way. The $N_{eff}$ for MPPC S13360-1325PE is found to be consistent with the real number of pixels of 2668\cite{sascha}.  Alternatively, $N_{eff}$ has been extracted using the AHCAL measurements with the SiPM mounted on a tile, showing consistent results. In this case, also saturation effects in the ASIC can play a role. For the reconstruction of hit energies from the SiPM signals the inverse of the saturation function is used.

\subsection{Time Calibration}

\begin{figure}[htp]
\centering
\includegraphics[width=8cm]{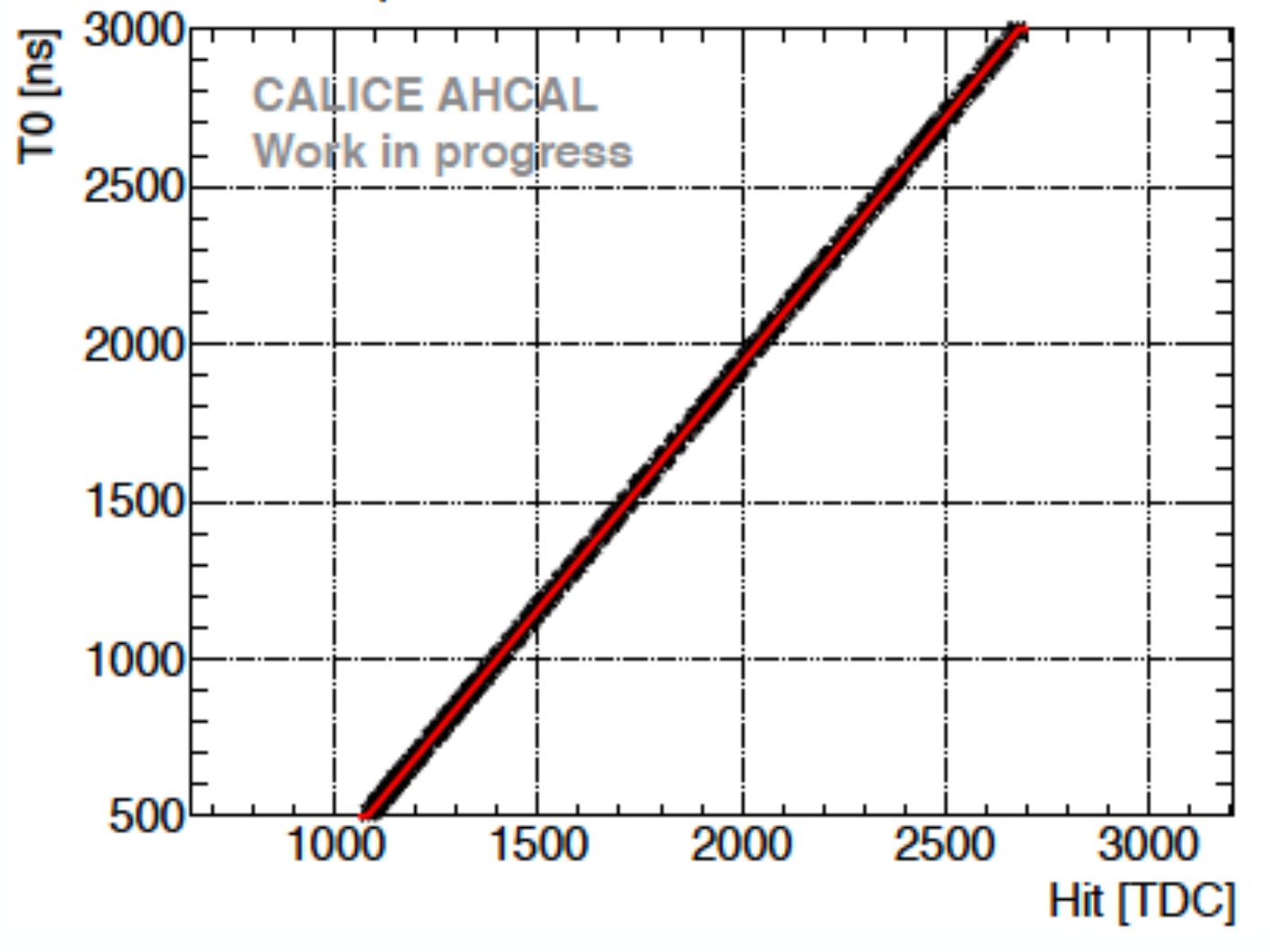}
\caption{TDC vs. time reference plot.}
\label{fig:time}
\end{figure}

The time calibration is performed using muon runs. The time information is stored in TDC values and needs to be calibrated to nanoseconds to have a common time reference in each event. To record time, TDC voltage is ramped up and down in intervals of 4000 ns. For this purpose the SPIROC2E chip has two TDC ramps for even and odd bunch crossings (bx) that form together a combined up and down ramping. Once a hit occurs the bxID as well as the current TDC value is recorded. From the slope of the TDC ramp and its pedestal the time in nanoseconds is calculated for both channel wise and memory cell-wise as shown in Fig \ref{fig:time}. A detailed description on how the time calibration is performed can be found in \cite{Ebra}. A 1 ns time resolution of the AHCAL is one of the design goal for testbeam mode. In practice the time resolution is limited by the front-end electronics.

\section{CONCLUSION}

The CALICE AHCAL is a SiPM-on-tile steel sampling calorimeter prototype which handles large number of SiPMs (\SI {}{\sim}22000 channels in total). It has been calibrated with an integrated LED system for SiPM gain, using minimum ionising particles (muons) for energy scale and corrected for the non-linearity of the SiPM response curves.  The detector has been successfully operated  in  test  beams from May to October 2018 at the CERN SPS and was stable with more than \SI{99.9}{\percent} channels working. With the calibrated prototype, shapes and structures of hadronic showers as well as the separation of two close-by showers can be studied in great detail, thus testing directly an important ingredient to the performance of particle flow algorithms. 

\section*{Acknowledgment}

I would like to thank my colleagues from AHCAL group for allowing me to present material and results from their research and colleagues from the CALICE collaboration for their help.



%





\end{document}